\documentclass[reprint,superscriptaddress,amsmath,amssym]{revtex4-2}

\usepackage{bm}
\usepackage{float}
\usepackage{graphicx}
\usepackage[usenames,dvipsnames]{color}
\usepackage[normalem]{ulem}
\usepackage[svgnames]{xcolor}
\usepackage{bm}
\usepackage{multirow}
\usepackage{titlesec}
\usepackage[utf8]{inputenc}
\usepackage{ragged2e} 
\begin{document}

\title{How heat propagates in liquid $^3$He}
\author{Kamran Behnia}
\affiliation{Laboratoire de Physique et d'\'Etude des Mat\'eriaux \\ 
(ESPCI - CNRS - Sorbonne Universit\'e), PSL Research University, 75005 Paris, France}
\author{ Kostya Trachenko}
\affiliation{School of Physical and Chemical Sciences, Queen Mary University of London, Mile End Road, London E1 4NS, United Kingdom
}
\date{\today}

\begin{abstract}
 In Landau's Fermi liquid picture, transport is governed by scattering between quasi-particles. The normal liquid $^3$He conforms to this picture but only at very low temperature. Here, we show that the deviation from the standard behavior is concomitant with the fermion-fermion scattering time falling below the Planckian time, $\frac{\hbar}{k_{\rm B}T}$ and the thermal diffusivity of this quantum liquid is bounded by a minimum set by fundamental physical constants and observed in classical liquids. This points to collective excitations (a sound mode) as carriers of heat. We propose that this mode has a wavevector of 2$k_F$ and a mean free path  equal to the de Broglie thermal length. This would provide an additional conducting channel with a $T^{1/2}$ temperature dependence, matching what is observed by experiments.  The experimental data from 0.007 K to 3 K can be accounted for, with a margin of 10\%, if thermal conductivity is the sum of two contributions: one by quasi-particles (varying as the inverse of temperature) and and another by sound (following the square root of temperature).
\end{abstract}

\maketitle

The impact of Landau’s Fermi liquid (FL) theory \cite{Landau1957} in condensed matter physics of the twentieth century can not be exaggerated. Before its formulation, the success of Sondheimer's picture of electrons in metals as a degenerate Fermi gas was a mystery \cite{giuliani_2008}. Even in a simple alkali metal such as Na, the Coulomb interaction between electrons is larger than the Fermi energy. Why wasn't this a problem for understanding the physics of metallic solids? Landau solved this mystery by proposing that a strongly interacting system of fermions can be mapped to an ideal system consisting of ``quasi-particles'' without interaction, or rather with an interaction weak enough to be considered as a perturbation \cite{nozieres1999theory,giuliani_2008}. His theory was inspired by the less common isotope of helium, namely $^3$He \cite{Dobbs},  the first experimental platform for testing the theory. Decades later, the theory was also applied to strongly correlated metals, known as heavy-fermion systems \cite{hewson_1993}.  

Leggett, reviewing  liquid $^3$He \cite{Leggett_2016}, writes that it is ``historically the first strongly interacting system of fermions of which we have been able to obtain a semi-quantitative description in the low-temperature limit.'' He also adds that the theory ``seems to agree quantitatively with experiment only for $T\lesssim$ 100 mK''. Compared to the degeneracy temperature of non-interacting fermions of the system ($\sim5$ K), this is a very low temperature. The properties of the ground state (and its evolution with pressure) have been the subject of numerous theoretical papers \cite{Abrikosov_1959,Brooker1968,Dy1969,wolfle1979,Vollhardt1984,calkoen1986,Vollhardt1987,Vollhardt1990}. In contrast, the breakdown of the Fermi liquid picture at very low temperatures, earlier noted by Emery \cite{EMERY19641} and Anderson \cite{Anderson1965}, ceased to be widely debated afterwards.

\begin{figure*}[ht]
\begin{center}
\includegraphics[width=15cm]{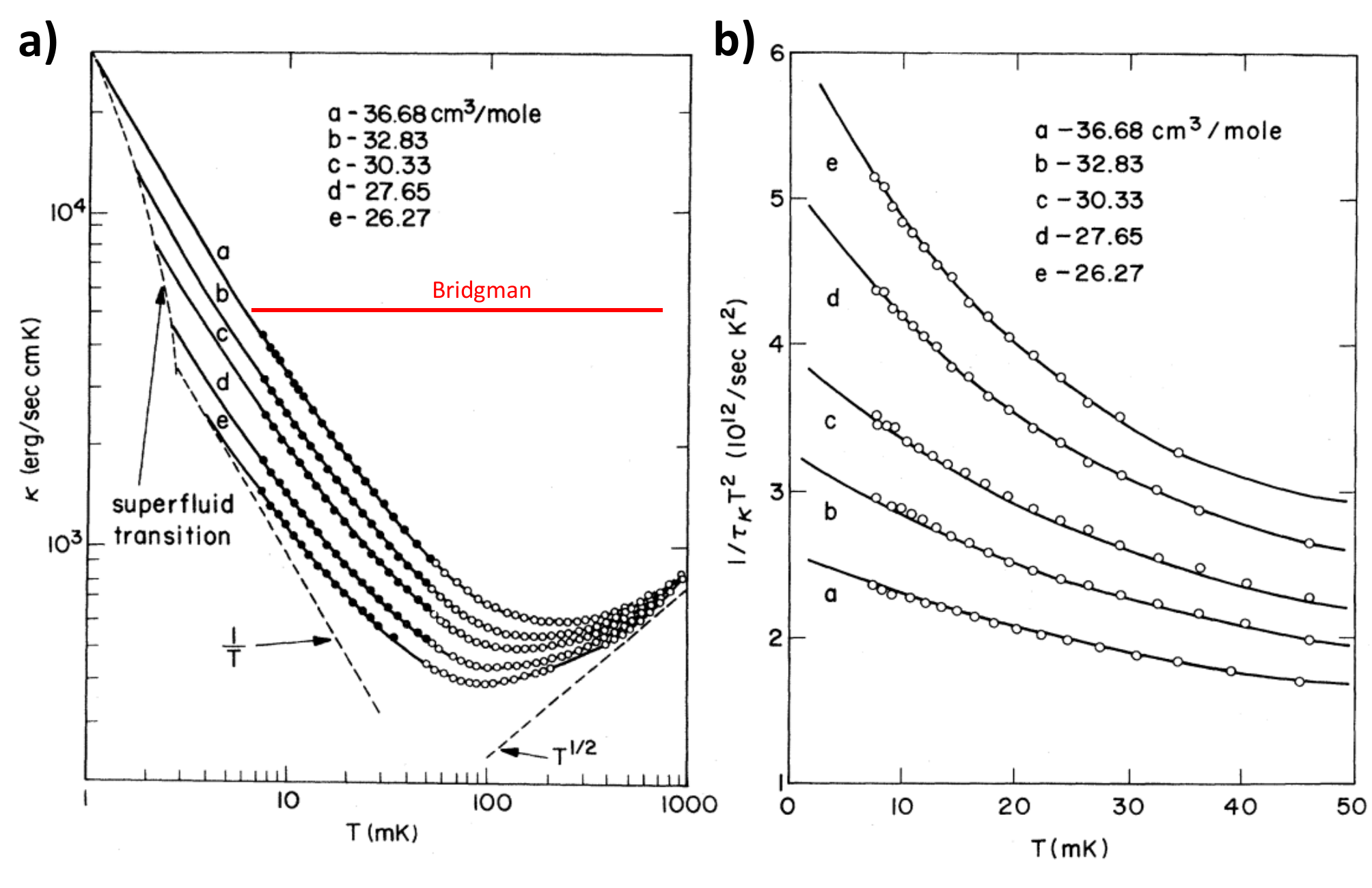}
\caption{\textbf{The narrow validity of the Fermi liquid picture of the thermal conductivity in $^3$He: }
a) Thermal conductivity, $\kappa$ as a function of temperature. b) The scattering time $\tau_{\kappa}$ extracted from the same data and specific heat, times the square of temperature, $T^2$. Contrary to what is expected in the standard Fermi liquid theory, $\tau_{\kappa}T^2$ is never constant. The figures are from Ref. \cite{greywall1984}. The horizontal red solid line in panel (a) represents what is expected for a classical liquid according to Bridgman's formula. Reproduced with the permission from the American Physical Society.}
\label{Fig-Greywall}
\end{center}
\end{figure*}

\begin{figure*}[ht]
\begin{center}
\includegraphics[width=17cm]{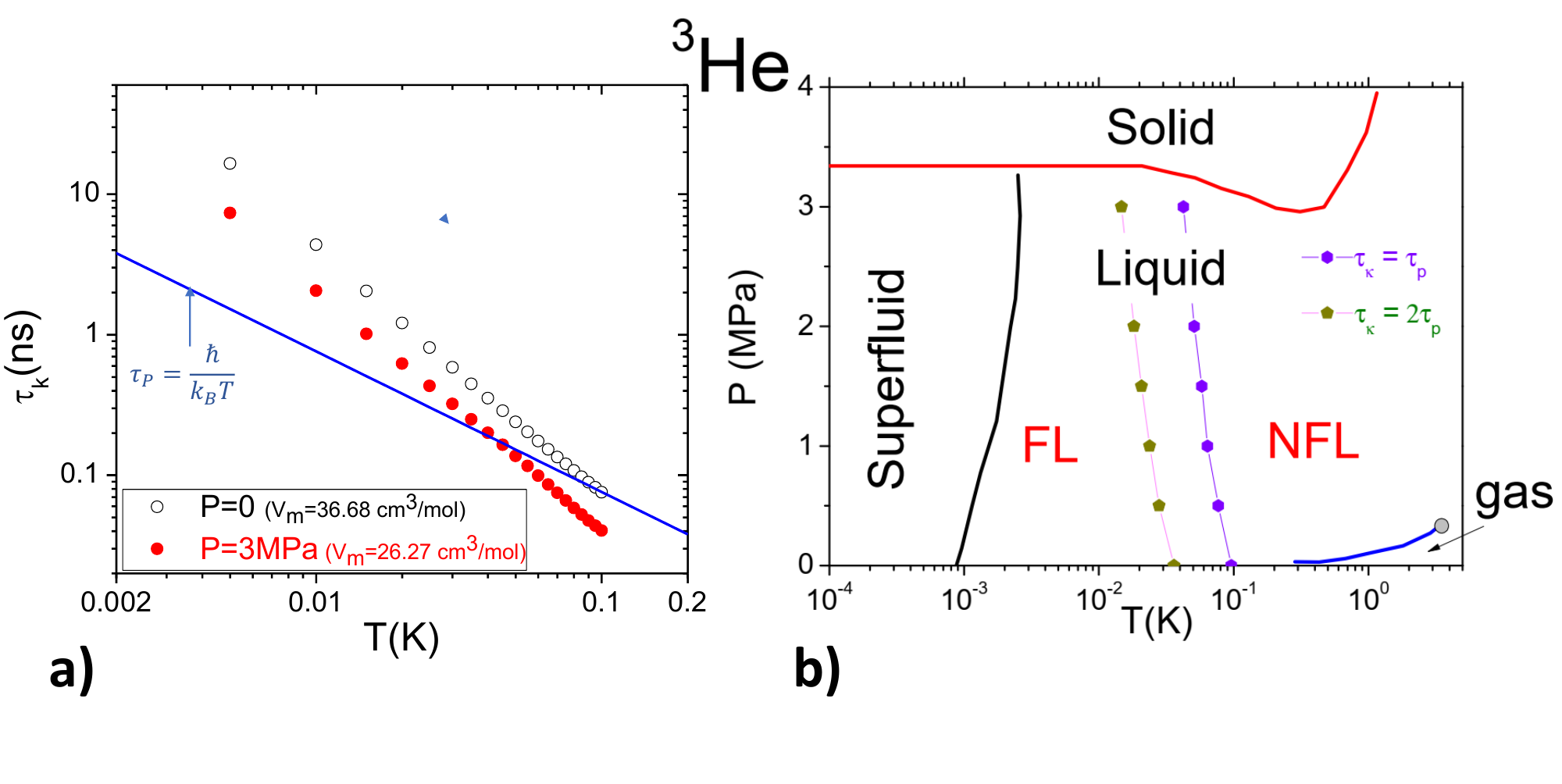}
\caption{\textbf{Scattering time, Planckian bound and the FL-NFL cross-over: } a) Fermion-fermion scattering time, $\tau_{\kappa}$ as a function of temperature  for two molar volumes \cite{greywall1984}. The blue solid line represents the Planckian time: $\tau_P=\frac{\hbar}{k_BT}$. Note that $\tau_{\kappa}$ tends to fall below $\tau_P$ at sufficiently high temperature. b) The phase diagram of $^3$He \cite{wolfle1979} and the fuzzy border between the Fermi liquid (FL) and the non-Fermi liquid (NFL) regimes. Also shown are temperatures below which $\tau_{\kappa}>\tau_P$ or $\tau_{\kappa}> 2\tau_P$.}
\label{Fig-tauP}
\end{center}
\end{figure*}

Here, we begin by recalling that as low as $T\approx 0.01$~K, the Fermi liquid picture does not hold. The Fermi temperature, with the mass normalization taken into account, is T$_F \sim 2$K. According to the most comprehensive set of data \cite{greywall1984}, when $\frac{T}{T_F} \approx 5\times 10^{-3}$ thermal conductivity, $\kappa$, deviates from the expected $T^{-1}$ behavior and the extracted scattering time, $\tau_{\kappa}$, is no more  $\propto T^{-2}$. We show that this ``non-Fermi-liquid'' (NFL) regime emerges when the fermion-fermion scattering time becomes comparable or shorter than the Planckian time \cite{Hartnoll2022,Bruin2013}, the time scale often invoked in the context of ``strange'' metallicity \cite{phillips2022}. Remarkably, in this regime, the thermal diffusivity, $D_{th}$, of quantum liquid $^3$He matches the minimum empirically observed \cite{Trachenko2021} and theoretically justified \cite{Trachenko2020,Trachenko2022} in classical liquids. We conclude that collective excitations play a role in heat transport comparable to the role played by phonons in classical liquids.  We find that the magnitude and the temperature dependence of the thermal conductivity can be accounted for if heat is carried by a phonon-like zero sound mode with a $2k_F$ wave-vector and an evanescence set by the thermal thickness of the Fermi surface in the momentum space \cite{Behnia2015b}. This phononic mechanism of heat propagation is distinguished from all those previously identified in solids and liquids, either classical or quantum. On the other hand, when the temperature becomes of the order of the Fermi temperature, its expression becomes another version of the Bridgman formula \cite{Brigman1923} for classical liquids. 

Fig. \ref{Fig-Greywall} reproduces figures reported by Greywall, who performed the most extensive study of thermal transport in normal liquid $^3$He \cite{greywall1984}. Samples with different molar volumes correspond to different pressures in the $T=0$ limit. As seen in Fig. \ref{Fig-Greywall}a, thermal conductivity, $\kappa$, at low temperature is inversely proportional to temperature, as expected in the FL picture. But the temperature window for this behavior, already narrow at zero pressure, shrinks with increasing pressure. By the melting pressure, the FL regime has almost vanished. The breakdown is even more visible in Fig. \ref{Fig-Greywall}b. It shows the temperature dependence of the inverse of of $\tau_{\kappa} T^2$, the scattering time extracted from thermal conductivity multiplied by the square of temperature, which should be constant in the Fermi liquid picture. A deviation is visible even at 8 mK and shoots up with increasing pressure. 

The deviation from the Fermi liquid behavior was usually attributed to spin fluctuations (see for example \cite{Dy1969}). While such a correction is expected at very low temperature, it is hard to see how they can play a role in our temperature of interest given the small amplitude of the exchange energy. (See the section 1 in the Supplement for details).




In Fig. \ref{Fig-tauP}a, we compare the temperature dependence of $\tau_{\kappa}$ according to Greywall's data (Fig. \ref{Fig-Greywall}b)) with the Planckian time, $\tau_P=\frac{\hbar}{k_BT}$ \cite{Hartnoll2022,Bruin2013}. One can see that, at zero pressure, $\tau_{\kappa}$ becomes of the order of $\tau_P$ at $T \approx 0.1$~K. At 3 MPa, near the melting pressure, $\tau_{\kappa}$ falls below $\tau_P$ at $\approx 0.043$~K. As seen in Fig. \ref{Fig-tauP}b, which reproduces the phase diagram of $^3$He \cite{wolfle1979, Dobbs,Vollhardt1990}, the crossover between the FL and the NFL regions of the phase diagram is concomitant with the passage from $\tau_\kappa \gg \tau_P$ to $\tau_\kappa  \lesssim \tau_P$. The possibility of a Planckian bound on dissipation is a subject  hotly debated in condensed matter physics \cite{Hartnoll2022}.

\begin{figure*}[ht]
\begin{center}
\includegraphics[width=17cm]{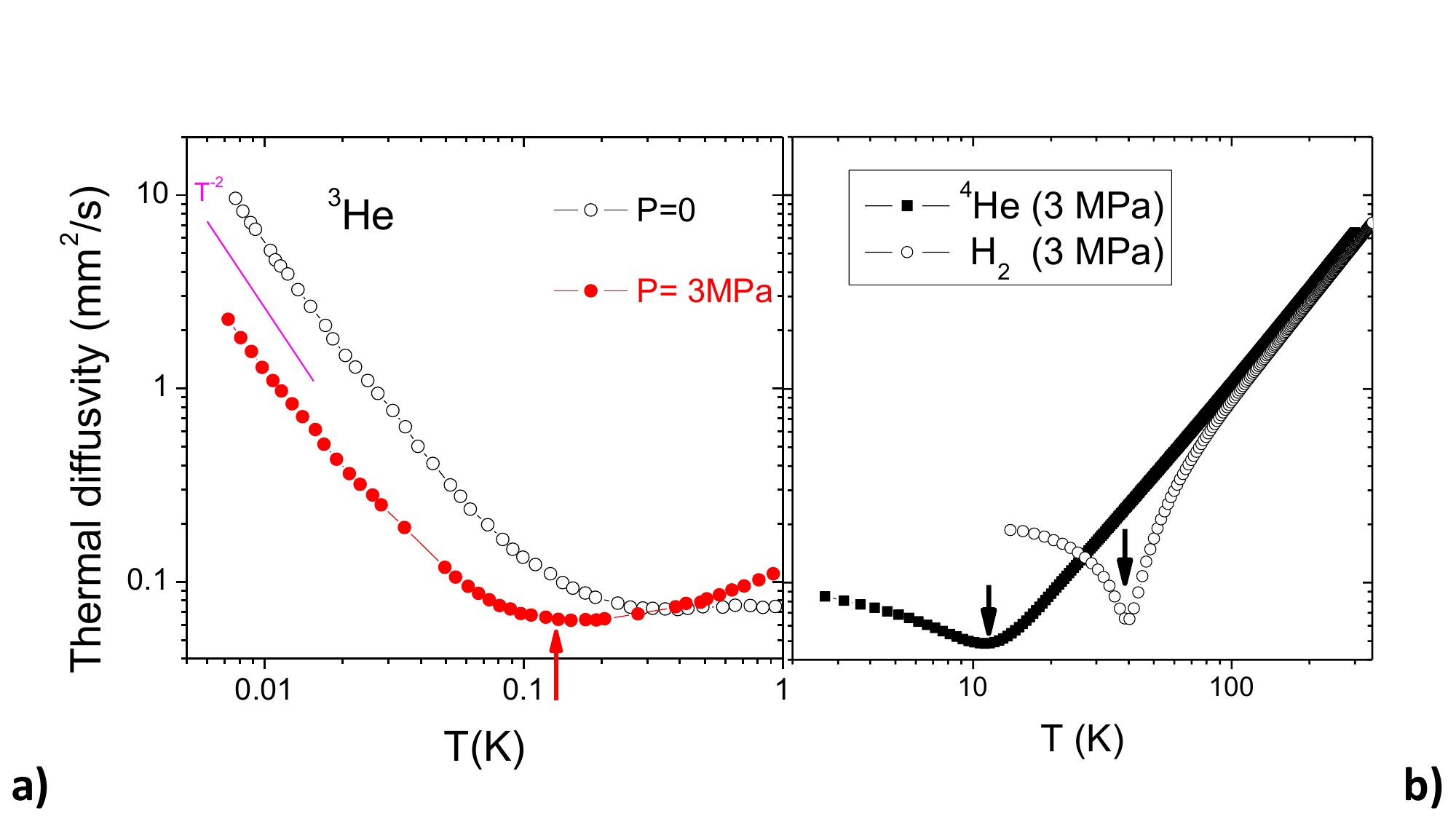}
\caption{\textbf{Bounds to thermal diffusivity:}
a) Thermal diffusivity, $D_{th}=\kappa/ C$,  of $^3$He, as a function of temperature. The Fermi liquid regime ($\propto$ $T^{-2}$) is restricted to low temperatures. It is followed by a saturation and a minimum. b) Thermal diffusivity as a function of temperature in two classical fluids (H$_2$ and $^4$He in the classical regime). In all three cases, $D_{th}$ has close values at the minimum, shown by the arrows.}
\label{Fig-D}
\end{center}
\end{figure*}

An important clue is provided by the temperature dependence of thermal diffusivity, $D_{th}$, obtained from $\kappa$ \cite{greywall1984} and specific heat \cite{greywall1983}. As seen in Fig. \ref{Fig-D}a, it shows a minimum,  both at zero pressure and at 3 MPa.

In all classical fluids, thermal diffusivity goes through a minimum at the intersection between a liquid-like regime, where it decreases with temperature and a gas-like regime where it increases with temperature \cite{nist,Trachenko2021}. We illustrate this in Fig. \ref{Fig-D}b, which shows the temperature dependence of $D_{th}$ in two fluids with slightly different atomic or molecular masses, namely H$_2$ and $^4$He \cite{nist}. In all three cases, there is a minimum of thermal diffusivity, consistent with other classical liquids \cite{nist,Trachenko2021} (See section 2 in the Supplement for more details). Although the minima are seen at different temperatures, the minimum $D_{th}$ has a similar amplitude: Expressed in mm$^2$s$^{-1}$, $D_{min}$, is $0.063$ in $^3$He, $\sim0.049$ in $^4$He and $\sim0.065$ in H$_2$. As discussed in the Supplement, this minimum is set by fundamental physical constants.

The closeness of the minima in $^3$He and other classical liquids suggests an important role played by phonon-like collective excitations of $^3$He in heat transport as they do in classical liquids considered earlier \cite{Trachenko2021}. Indeed, diffusion at high temperature is driven by the random walk of the particles. Cooling lowers the diffusion constant by decreasing the velocity and the mean free path close to interatomic separation. Below a given temperature, collective excitations (e.g., sound) begin to operate. In a quantum liquid, this process occurs along the opposite direction: the diffusion constant is dominated by quasi-particles at low temperature. With warming, the mean path decreases and approaches its minimal value as in classical liquids \cite{Trachenko2021}, as is seen in Fig. \ref{Fig-D}a. When thermal conductivity due to quasiparticles becomes small, the other remaining mechanism becomes important, namely conductivity due to collective excitations, sound. Using Landau's own words,``sonic excitations in the gas of quasi-particles (phonons of the ``zeroth sound'')'' \cite{Landau1959}, are to become the main carriers of heat above this minimum.

A collision time shorter than the Planckian time means that the frequency of thermally excited zero sound (which increases linearly with temperature ($\omega_{zs}=\frac{k_BT}{\hbar})$) becomes smaller than the scattering rate (which increases quadratically with temperature).  This inequality ($\omega_{zs} \tau_{\kappa} < 1 $) means that the thermally excited zero sound is in the hydrodynamic limit, where the distinction between zero sound and first sound fades away \cite{nozieres1999theory}. In liquid $^3$He, this occurs at a remarkably low temperature, because  $\tau_{\kappa}T^2 \ll \frac{\hbar E_F}{k_B^2}$ \cite{Behnia2022b} (see section 3 in the Supplement).

\begin{figure}[ht]
\includegraphics[width=9cm]{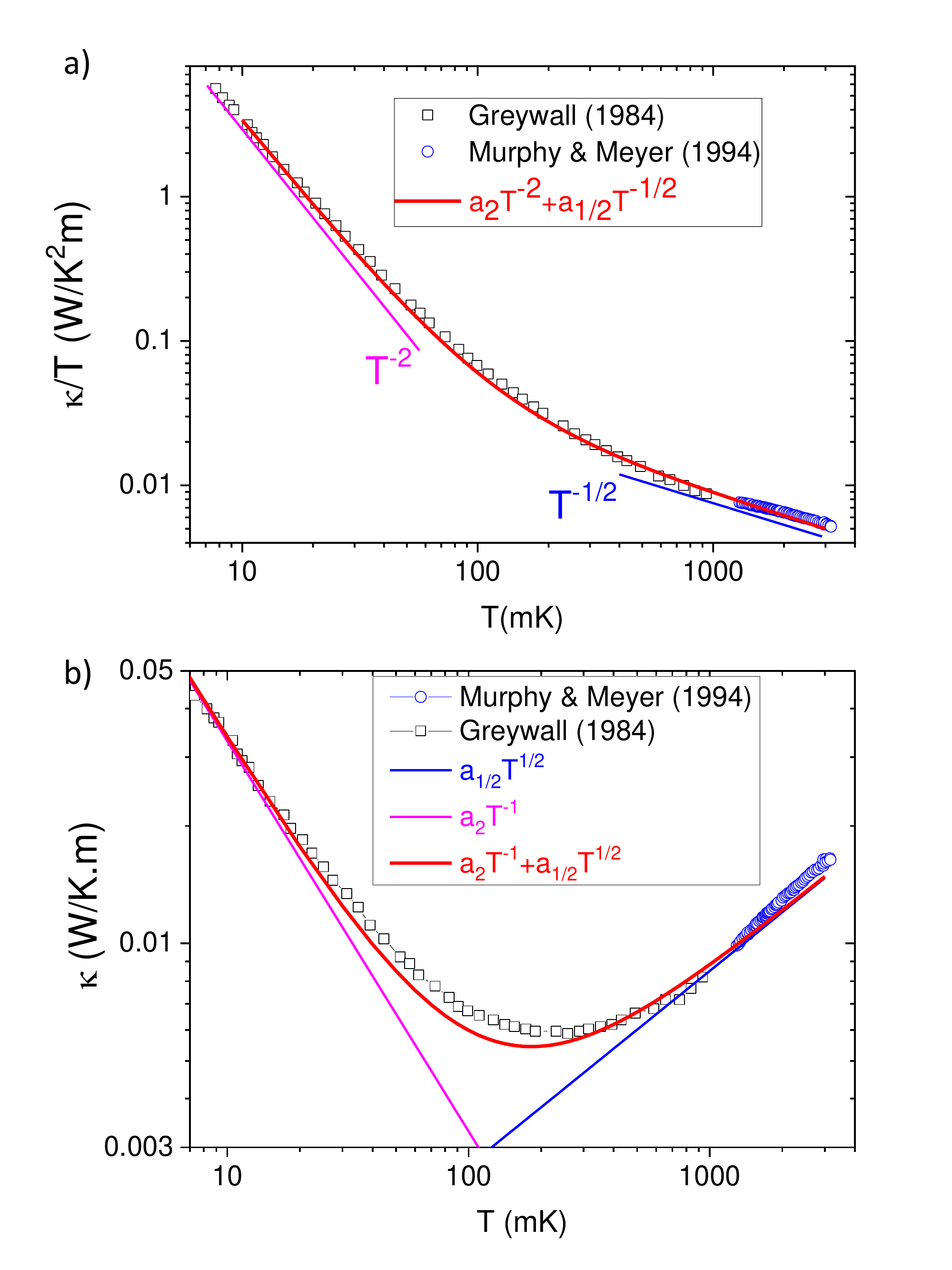}
\caption{\textbf{Experimental data and our model.} a) Thermal conductivity divided by temperature as reported by Greywall \cite{greywall1984} (at zero pressure) and by Murphy and Meyer \cite{Murphy1994} (at saturated vapor pressure). At very low temperatures ($T\sim 0.01$ K) in the Fermi liquid regime, $\kappa/T \propto T^{-2}$ (purple line). Above 0.5 K, $\kappa/T \propto T^{-1/2}$. The blue line represents what is expected by Eq. \eqref{kappa_zs}, using the effective mass and the carrier density of $^3$He. The red solid line represents a fit to the experimental data in the whole temperature range assuming that $\kappa/T$ consists of the sum of  $T^{-1/2}$ (sound transmission) and  $T^{-2}$ term (quasi-particle transmission) terms. b) Same data plotted for $\kappa(T)$. The maximum discrepancy between data and theory is about 10\%. Also shown are the two components of the total $\kappa(T)$.}
\label{Fig-2-regimes}
\end{figure}

As seen in Fig. \ref{Fig-Greywall}a, Greywall observed that above 0.5 K, $\kappa \propto T^{1/2}$. Fig. \ref{Fig-2-regimes}a shows the temperature dependence of $\kappa/T$ in normal liquid $^3$He. It includes Greywall's data at $T<1$ K \cite{greywall1984}, measured at constant molar volume of 36.68 cm$^3$/mol (corresponding to zero pressure in the low temperature limit) and the data reported by Murphy and Meyer \cite{Murphy1994}, measured at saturating vapor pressure (SVP) above 1.2 K (Fig. \ref{Fig-2-regimes}a). Despite the imperfect agreement (unsurprising given the change in the molar volume at SVP), one can see that Murphy and Meyer \cite{Murphy1994} roughly confirm Greywall's observation about the asymptotic tendency of thermal conductivity: $\kappa\propto T^{1/2}$. This temperature dependence is distinct from what is known to occur in different regimes of phonon thermal conductivity in crystals and glasses (see Table \ref{Table_ph}) and also from the Bridgman formula ($\kappa_{B}=r k_Bn^{2/3}v_s$) \cite{Brigman1923,bird2002transport,Zhao2021, Xi_2020} for classical liquids, which does not contain any temperature dependence.

\begin{table*}[ht!]
\centering
\begin{tabular}{|c|c|c|c|c|c|c|}
\hline
System & Sonic heat carriers & Temperature dependence & Mechanism & reference\\
\hline
\hline
Crystal  ($T\rightarrow 0$) & small-$q$ phonons &$\kappa \propto T^{3}$& Boundary scattering &  \cite{Berman1976,ashcroft2011solid}\\
\hline
Crystal ($T\sim T_{\rm D}$) & large-$q$ phonons  &$\kappa \propto T^{-1}$& Umklapp scattering &  \cite{Berman1976,Behnia_2019,Mousatov2020}\\
\hline
Glass ($T\rightarrow 0$) & ``propagons'' &$\kappa \propto T^{\sim2}$ & Rayleigh scattering &  \cite{Zeller1971,Allen1989,Allen1999}\\
\hline
Glass (high T) & ``diffusons'' &$\kappa \propto T^{0}$ & Minimum mean free path &  \cite{Cahill_1992,Allen1999,Xie_2017}\\
\hline
Classical liquid  &large-$q$ phonons  &$\kappa \propto T^{0}$ & Minimum mean free path&  \cite{Brigman1923}\\
\hline
Quantum liquid $^3$He  & $q\sim2k_F$ phonons &$\kappa \propto T^{1/2}$ & Fermi surface thermal fuzziness&  This work \\
\hline
\end{tabular}
\caption{\textbf{Different cases of phonon thermal conductivity:} Comparison between the present case of heat propagation by collective excitations with other and better-understood regimes of phonon thermal conductivity.}
\label{Table_ph}
\end{table*}

Here, $\frac{\pi^2}{3}\frac{k_B^2}{h}$ is the quantum of thermal conductance \cite{Schwab2000}. The transmission coefficient $\mathcal{T}$ is set by the number of carrier modes and their mean free path. Its units depends on dimensions (meters in 1D, dimensionless in 2D and meters$^{-1}$ in 3D). 

In three dimensions, with a spherical Fermi surface of radius $k_F$, there are $\frac{8\pi}{3\lambda_F^2}$ conducting modes and with mean-free-path set by fermion-fermion scattering,  $\ell_{ff}$, the thermal conductivity becomes :

\begin{equation}
\frac{\kappa}{T}|_{qp}=\frac{2\pi}{9}\frac{k_B^2}{h} k_F^2\ell_{ff}
\label{qp}
\end{equation}

\begin{figure*}[ht]
\begin{center}
\includegraphics[width=16cm]{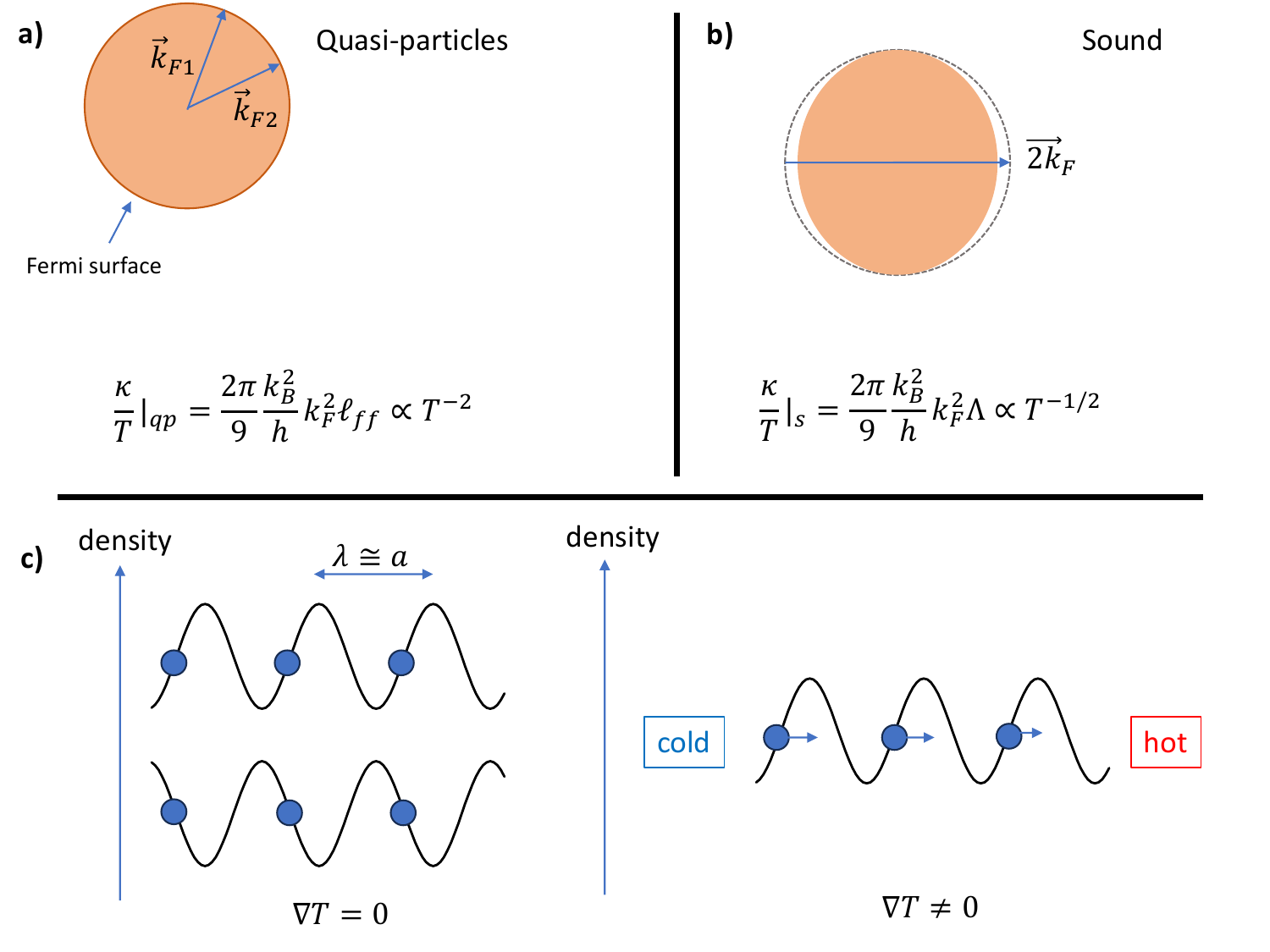}
\caption{\textbf{Two channels for the conduction of heat: }
a) Transmission by quasi-particles is dominant at low temperatures. Collisions in momentum space between quasi-particles lead to a $\propto T^{-2}$ transmission. b) Transmission by collective response of the Fermi surface. The amplitude of transmission, set by the square of the Fermi radius and the de Broglie thermal wavelength, the thermal thickness of the Fermi surface, is $\propto T^{-1/2}$. c) The sound with a 2$k_F$ wavevector has a wavelength of the order of the interatomic distance. A thermal gradient will cancel the equivalency between two modes with a $\pi/2$ phase shift.}
\label{Fig-2-channels}
\end{center}
\end{figure*}

At very low temperature, the response to temperature gradient is dominated by quasi-particles (Fig. \ref{Fig-2-regimes}b). These are plane waves within a thermal window of the Fermi level which can carry  heat. $\kappa/T$  decreases quadratically with temperature, due to the temperature dependence of the mean free path of quasi-particles, $\ell_{qp}$. 

In order to identify the collective transport mode leading to the $T^{1/2}$ temperature dependence dominant above 0.5 K, let us compare it with two other cases. In crystals, the cubic temperature dependence of phonons at low temperature reflects the temperature dependence of the volume of the Debye sphere when the mean-free-path is saturated to a constant a value. In glasses, the asymptotic low temperature dependence of thermal conductivity by ``propagons'' is  close to quadratic. The slower temperature dependence  $\kappa$ is due to an increase in mean-free-path  with cooling. In both cases, the presence of long wavelength carriers leads to a superlinear exponent (between 2 and 3) in the temperature dependence. Our case requires a scenario circumventing the cubic temperature dependence of a Debye sphere. 

A collective transmission by the whole Fermi surface will meet this requirement. This would be a sound mode with a wave-vector fixed at twice the  Fermi radius (Fig. \ref{Fig-2-regimes}c). There are two reasons for distinguishing $2k_F$ as a wavevector. The first is theoretical. The Lindhard function, which quantifies the susceptibility of a fermionic gas to an external perturbation has a singularity at $q=2k_F$ \cite{ashcroft2011solid}. The second is experimental. Inelastic X-ray scattering experiments \cite{Albergamo2007,Albergamo2008} find that the dispersion of zero sound has a pronounced anomaly near $q=2k_F$ (see section 4 in the Supplement). 

The Landauer transmission rate of such a heat-carrying mode depends on its mean free path. The wave is attenuated by the thermal fuzziness of the Fermi surface in the momentum space, which is set by the inverse of the   the de Broglie thermal length \cite{Behnia2015b}. Therefore, we can take the mean free path to be $\Lambda = \frac{h}{\sqrt{2\pi m^* k_B T}}$, ($m^*$ is the effective mass). The number of involved states is identical to the one used to quantify the quasi-particle contribution. Replacing $\mathcal{T}$ in Eq. \eqref{Landauer} then gives:
\begin{equation}
\frac{\kappa}{T}|_{s}=\frac{2\pi}{9}\frac{k_B^2}{h} k_F^2 \Lambda
\label{zs}
\end{equation}

Substituting $\Lambda$ by its explicit value and the Fermi wave-vector with the particle density (with $n=\frac{k_F^3}{3 \pi^2}$) leads to:

\begin{equation}
\kappa_{s}\simeq r k_B n^{2/3}\sqrt{\frac{2 k_B T}{m^*}}
\label{kappa_zs}
\end{equation}

Here  $r=\frac{\pi^{11/6}}{ 3^{4/3}}\approx 1.88$. 

Equation \eqref{kappa_zs} has two parameters, the particle density, $n$, and the effective mass, $m^*$. In Fig. \ref{Fig-2-regimes}a, it is plotted using the effective mass ($m^*$=2.7 $m_3$ \cite{greywall1983}= 1.35 $\times 10^{-26}$ kg) and the zero-pressure carrier density  ($n=1.64\times 10^{28}$ m$^3$, corresponding to a molar volume of 36.68 cm$^3$/mol \cite{greywall1984}) of normal liquid $^3$He. At  1 (2) K, the experimentally measured thermal conductivity is 11 (20) percent larger what is expected from equation $\eqref{kappa_zs}$. 

If there is an additional conduction channel by sound then, the narrow validity of the standard Fermi liquid approach, finds an explanation too. As the red line in Fig. \ref{Fig-2-regimes}a shows, at any arbitrary temperature between 0.01 K and 3 K, the experimentally measured $\kappa/T$ can be described by a sum of $T^{-2}$ and $T^{-1/2}$ terms. Fig. \ref{Fig-2-regimes}b shows the same data in a plot of $\kappa(T)$. One can see that in the whole temperature range, the total thermal conductivity can be expressed as a sum of two terms:

\begin{equation}
\kappa(T)= \kappa_{qp}+\kappa_{s}
\label{total}
\end{equation}

\noindent where $\kappa_{qp}\propto T^{-1}$ is given by Eq. \ref{qp} and $\kappa_{s}\propto T^{1/2}$ is given by  Eq. \eqref{zs} (or equivalently by Eq. \eqref{kappa_zs}).

\begin{table*}[ht!]
\centering
\begin{tabular}{|c|c|c|c|}
\hline
 a$_2$ (10$^{-4}$W$\cdot$ m$^{-1}$) &  a$_{1/2}$ 
 (10$^{-2}$ W$\cdot$m$^{-1}\cdot$ K$^{-3/2}$) &  Reference\\
\hline
\hline
3.3 &  8.5 &  This work (Fit in Fig. \ref{Fig-2-regimes})\\
 \hline
-- &  7.6 &  This work (Eq. \ref{kappa_zs}:  $m^*=2.7m_3$ \& $n=1.63\times 10^{22}$)\\ 
\hline
2.9 &  -- &  Ref. \cite{greywall1984} (experiment)\\ 
\hline
3.5 &  -- &  Ref. \cite{Wheatley1968} (experiment)\\ 
\hline
3.3-5.4 &  -- &  Ref. \cite{Dy1969} (theory)\\ 
\hline
5 &  -- &  Ref. \cite{Brooker1968} (theory)\\ 
\hline
\end{tabular}
\caption{\textbf{Numerical amplitudes of a$_2$ and a$_{1/2}$:} 
 The parameters of the  fit to $a_2T^{-2}+a_{1/2}T^{-1/2}$ plotted in Fig. \ref{Fig-2-regimes}. Also listed are a$_{1/2}$ according to Eq. \eqref{kappa_zs} (with no adjustable parameter) and previous experimental and theoretical reports on the magnitude of the prefactor of the quasi-particle contribution, $a_2[\equiv \kappa T]$.}
\label{Table_coeff}
\end{table*}

The fit parameters used for the red curves in Fig. \ref{Fig-2-regimes},  a$_2$ and a$_{1/2}$ for $\kappa_{qp}/T=a_2T^{-2}$ and $\kappa_{s}/T=a_2T^{-1/2}$, are listed in table \ref{Table_coeff}. The table also compares the amplitude of $a_2$  with previous  experimental \cite{greywall1984,Wheatley1968} and theoretical \cite{Brooker1968,Dy1969} estimations of it.

Given the simplicity of the picture drawn above, this is a surprisingly good agreement. Let us recall that the in the case of the S.V.P. data by Murphy and Meyer \cite{Murphy1994}, particle density is not constant and decreases with warming. Moreover, the effective mass also changes with temperature. Finally, not only our simple model neglects any change in density and mass, it also does not take in to account a finite coupling between the two carriers of heat (particles and sound). Nevertheless, the quantitative difference between the expected and the measured thermal conductivity does not exceed 10 percent. 

Eq. \eqref{kappa_zs} has a striking resemblance to the Bridgman formula for classical liquids \cite{Brigman1923,bird2002transport,Zhao2021,Xi_2020} (see Supplementary note 6), with the speed of sound replaced by a group velocity of $\sqrt{\frac{2k_BT}{m}}$. This can be accounted for by noticing that the time scale for randomness in a quantum liquid is set by  the ratio of the Fermi velocity to the de Broglie thermal length (and not by the ratio of the sound velocity to the inter-particle distance, as in a classical liquid or a glass). Interestingly, the two equations become similar at the classical/quantum boundary, that is when the de Broglie thermal length becomes equal to the Fermi wavelength (see Supplementary note 6). 

Fig. \ref{Fig-2-channels} summarizes our main message. It appears that in a Fermi liquid two modes of conduction are at work. The first (Fig. \ref{Fig-2-channels}a) has been understood for decades and is based on collisions between quasi-particles. The second, identified here, is the breathing of the whole Fermi surface  (Fig. \ref{Fig-2-channels}b). Interestingly, the two channels differ only by their respective relevant length scale, the distance between two successive collisions and the thermal de Broglie length.

In the real space, this collective mode is presumably a visco-elastic \cite{Rudnick1980,Dobbs} soft phonon with the shortest possible wavelength, which is the interatomic distance. As seen in Fig. \ref{Fig-2-channels}c, the wavelength of the sound in question is almost identical to interatomic distance. Such a wave can be generated either by a leftward or rightward shift of all atoms. A thermal gradient lifts this degeneracy. The fundamental reason behind the success of this simple approach is yet t be rigorously understood. 

After this paper was written, we learned of two relevant early works, by Brazovskii \cite{Brazovskii} on the role played by a soft mode in the crystallization of a liquid, and by Dyugaev \cite{Dyugaev} on 2k$_F$ rotons in $^3$He. 

Arguably, normal liquid $^3$He is the cleanest and the simplest known Fermi liquid. If collective excitations play such a central role in its transport properties across such a wide temperature range, what about other strongly interacting systems of fermions residing beyond the Landau's paradigm \cite{Keimer2015}? We leave this question to future studies.

\section*{Acknowledgements} K. B. acknowledges discussions with Mikhail Feigelman and is supported by the Agence Nationale de la Recherche (ANR-19-CE30-0014-04) and by the National Science Foundation (under Grant No. NSF PHY11-25915). K. T. is grateful to EPSRC for support.

\section*{Author Contributions} The main idea of the paper was born during the discussions between the two authors. K.B wrote the text with continuous feedback from K. T. .

\section*{Competing Interests} The authors declare to have no competing interests. 

\bibliography{biblio}
\clearpage
\renewcommand{\thesection}{S\arabic{section}}
\renewcommand{\thetable}{S\arabic{table}}
\renewcommand{\thefigure}{S\arabic{figure}}
\renewcommand{\theequation}{S\arabic{equation}}
\setcounter{section}{0}
\setcounter{figure}{0}
\setcounter{table}{0}
\setcounter{equation}{0}

\begin{widetext}

\begin{center}{\large\bf Supplementary Materials for \emph{How heat propagates in non-Fermi liquid $^3$He}}\\
\end{center}

\subsection*{Supplementary Note 1: Departure from Fermi liquid behavior due to spin fluctuations}
Several early attempts to explain the non-Fermi liquid behavior observed in $^3$He invoked spin fluctuations.
Since the exchange energy of spins is in the range $\approx 1$ mK, this approach is not sufficient to explain the non-Fermi liquid behavior seen in a temperature range, which is at least two orders of magnitude larger. Nuclear magnetic relaxation measurements have found that the exchange energy of spins is $J$=0.002 K when the molar volume is 24.6 cm$^3$/mol \cite{Meyer1968}. Specific heat measurements find that $J$=0.00085 K, when the molar volume is 24.45 cm$^3$/mol \cite{Greywall1977}. These numbers are to be compared to the Fermi ($\approx 2$ K) and Debye energy ($\approx 20$ K). There is no detectable evidence of a role played by spin fluctuations  in heat transport (either as conductors of heat or as scattering centers) when the temperature exceeds 0.1 K. 

\subsection*{Supplementary Note 2: Lower bounds on thermal diffusivity and viscosity in classical liquids}

In liquid and supercritical states of matter, thermal conductivity and thermal diffusivity are strongly system-dependent and can change by many orders of magnitude depending on temperature. However, there is a universal theoretical lower bound of thermal diffusivity, $\alpha_{min}$, defined by \cite{Trachenko2021}:  

\begin{equation}
\alpha_{min}=\frac{1}{4\pi}\frac{\hbar}{\sqrt{m_e m}} 
\label{min}
\end{equation}

\noindent where $m_e$ is the electron mass and $m$ is the molecule mass.

\begin{figure*}[ht]

\includegraphics[width=14cm]{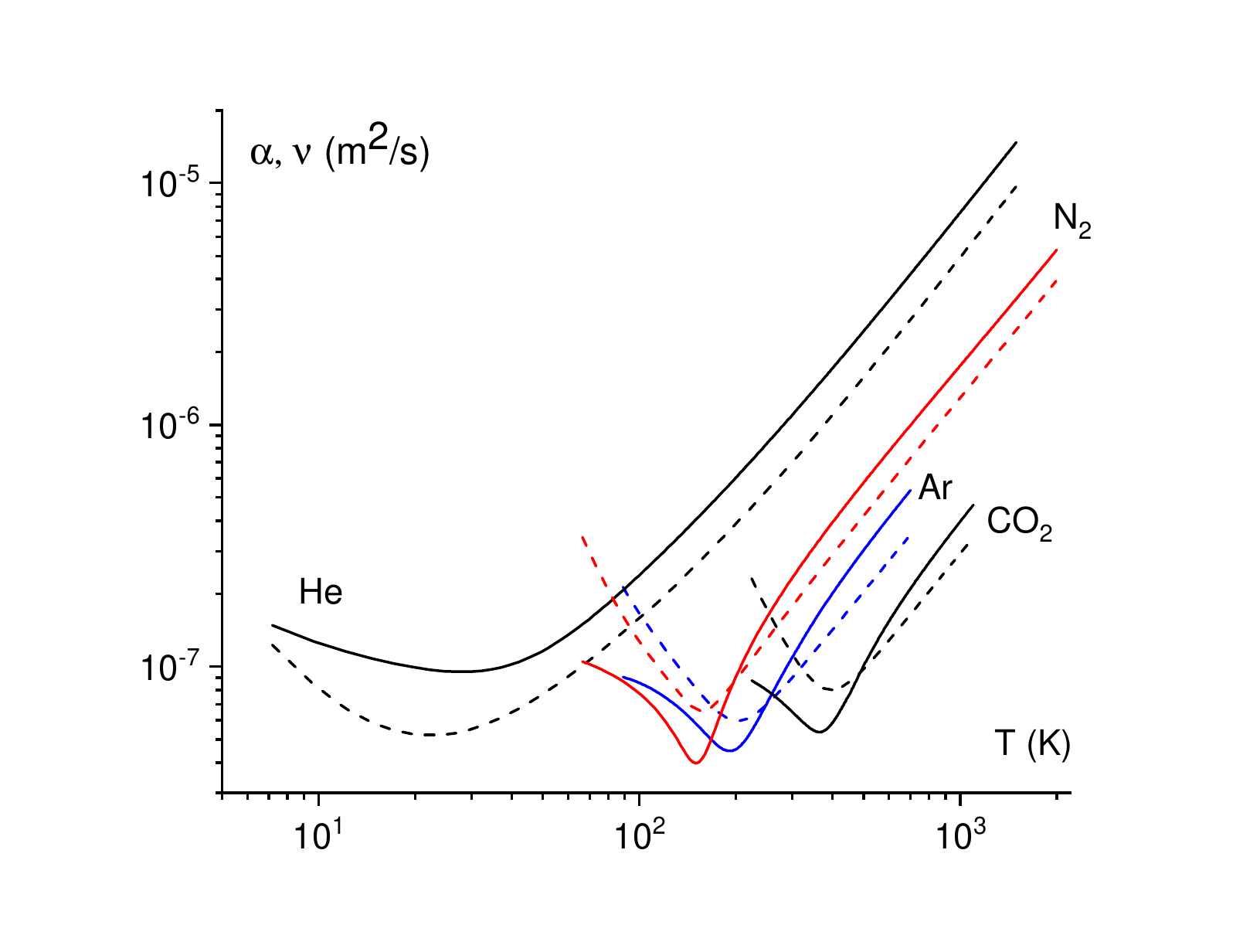}\\
\justifying{\textbf{Supplementary Figure 1: Minima of thermal diffusivity and kinematic viscosity in classical liquids.}
 Experimental curves of thermal diffusivity (solid lines) and kinematic viscosity (dashed lines) for $^4$He (20 MPa), N$_2$ (10 MPa), Ar (20 MPa), and CO$_2$ (30 MPa). }.
\label{Fig-universal}
\end{figure*}

The lower bound corresponds to the minima seen in the Supplementary Figure 1. The minimum itself is due to the dynamical crossover between the low-temperature liquidlike regime where each particle undergoes both oscillatory and diffusive motion and the high-temperature gaslike regime where the oscillatory component of motion is lost and only the diffusive component is present \cite{Trachenko2020,Trachenko2021,Trachenko2022}. The value of thermal diffusivity at the minimum can be evaluated by equating either the phonon mean free path in the liquidlike regime or the particle mean free path in the gaslike regime to the interatomic separation. Then, the value at the minimum is related to only two parameters in condensed matter: interatomic separation and Debye vibration frequency. Relating this length and energy scale to the fundamental physical constants using Bohr radius and Rydberg energy gives Eq. \eqref{min} \cite{Trachenko2021,Trachenko2022}.

This theoretical minimum in thermal diffusivity, $\alpha_{min}$, coincides with a minimum in the kinematic viscosity $\nu_{min}$ \cite{Trachenko2020}. As seen in Figure 1 of the main text, the experimental data confirms that the two minima are close to each other.

\begin{figure*}[ht]

\includegraphics[width=13cm]{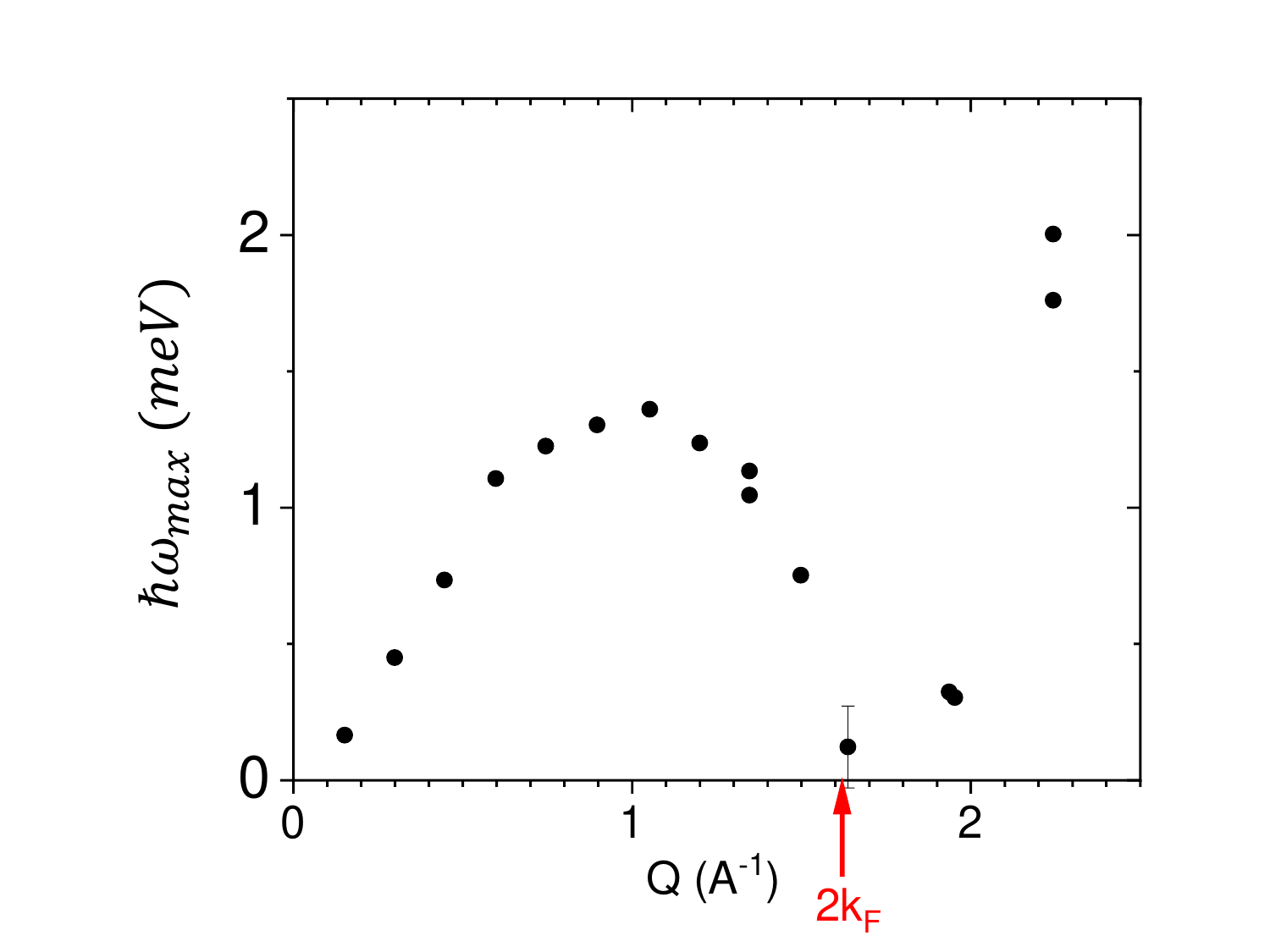}\\
\justifying{\textbf{Supplementary Figure 2: Phonon dispersion according to inelastic X-ray scattering experiments.} Acoustic dispersion curves for normal liquid $^3$He at 1.1 K \cite{Albergamo2007,Albergamo2008}. The red arrow points to $Q=2k_F$=1.59\AA$^{-1}$.   }
\label{Fig-SD}

\end{figure*}

\subsection*{Supplementary Note 3: The dimensionless fermion-fermion collision cross section }
The relevance of Planckian bound when the temperature is more than one order of magnitude below the degeneracy temperature of fermionic quasi-particles is driven by the unusually large fermion-fermion scattering rate in $^3$He \cite{Behnia2022b}. 

The dimensionless cross section of fermion-fermion scattering, $\zeta=\frac{\hbar E_F}{\tau_{\kappa}(k_BT)^2}$, in $^3$He is as large as $\sim 35$  at zero pressure and shoots up to  $\sim 60$ at the melting pressure. In metals, this cross section is often smaller \cite{Behnia2022}. When electrons are weakly scattered, $\zeta$ seldom exceeds unity. But in strongly correlated metals, it is as large as 10. In La$_{1.67}$Sr$_{0.33}$CuO$_4$ cuprate, $\zeta$ is almost as large as in $^3$He \cite{Behnia2022b}. 

\subsection*{Supplementary Note 4: Thermal conductivity in the Landauer picture}
Landauer pictured conduction as the transmission of a wave attenuated over a distance. When the wave is a ballistic quasi-particle, the attenuation distance (or the mean-free-path) is set by the finite size. One virtue of this approach is the transparency of the presence of fundamental constants and  context-dependent length scales in setting the expected magnitude and temperature dependence of a transport coefficient.

The Landauer formalism leads to the following expression for the electrical conductivity of a two-dimensional metal whose Fermi surface is a circle of radius $k_F$ and its mean-free-path $\ell$:
\begin{equation}
\sigma^{2D}=\frac{e^2}{h} k_F \ell
\end{equation}

This is because the number of conducting mode is $2\times1/2\times \frac{2\pi}{\lambda_F}=k_F$ (2 for spin degeneracy, $\frac{1}{2}$ for averaging a vector in two dimensions, and $2\pi$ representing the total planar angular range).  

In three dimensions, for a spherical Fermi surface  of radius k$_F$, there are $2\times 1/3\times \frac{4\pi}{\lambda_F^2}=\frac{2}{3\pi}k_F^2$ conducting modes. Here, 2 is for spin degeneracy, $\frac{1}{3}$ is for averaging a vector in three dimensions, and $4\pi$ represents the solid angle. Therefore: 

\begin{equation}
\sigma^{3D}=\frac{2}{3\pi}\frac{e^2}{h} k_F^2 \ell
\end{equation}

Note that both these expressions for electrical conductivity are strictly equivalent to the Drude formula: $\sigma= \frac{ne^2\tau}{m}$. 

To find an equivalent expression for thermal conductance, it is sufficient to multiply the expression for $\sigma^{3D}$  by the Sommerfeld value $L_0=\frac{\pi^2}{3}\frac{k_B^2}{e^2}$ leads to :

\begin{equation}
\frac{\kappa^{3D}}{T}=\frac{2\pi}{9}\frac{k_B^2}{h} k_F^2 \ell
\end{equation}

This is the equation 2 of the main text.

\subsection*{Supplementary Note 5: Phonon dispersion in normal liquid $^3$He.}
Albergamo \textit{et al.} used inelastic X-ray scattering to study the elementary excitations of normal liquid $^3$He at 1.1 K. They found that the zero sound wave remains well defined in the entire range of explored wave numbers (0.15$\leq$ $Q$ $\leq$ 3.15 \AA$^{-1}$) \cite{Albergamo2007,Albergamo2008}. 

Fig. \ref{Fig-SD} shows the dispersion curve taking into account the attenuation of the wave. One can see that the energy of the sound mode shows a deep minimum near $Q=2k_F$ \cite{Schmets2008,Albergamo2008}.

Krotscheck and Panholzer \cite{Krotscheck2011} performed a theoretical analysis of this data and found that ``pair fluctuations'' sharpen this mode in contrast to the random phase approximation (RPA). According to their conclusion, the collective mode has a dispersion similar to the phonon-maxon-roton dispersion of $^4$He. The minimum was found to be  close to the experimental value of 1.6 \AA$^{-1} $.


\subsection*{Supplementary Note 6: The Bridgman formula and the thermal conductivity in classical liquids, glasses and quantum liquids}
The Bridgman formula was written a century ago \cite{Brigman1923}. It relates the thermal conductivity to the sound velocity and particle concentration of dense liquids: 

\begin{equation}
\kappa=r k_{\rm B} n^{2/3}  v_s 
\label{bridgman}
\end{equation}

Different authors assume either $r=3$ \cite{bird2002transport} or $r=2$ \cite{Zhao2021,Xi_2020}. The specific heat of liquids at low temperature is close to the Dulong-Petit value, 3$k{\rm _B}$, and decreases to 2$k{\rm_B}$ at high temperature \cite{ropp}, corresponding to $r=3$ and $r=2$, respectively. The Bridgman formula yields a  surprisingly good account of available data in liquids. The thermal conductivity of water is in agreement with $r\simeq 3$. For ethanol, methanol, butane, and pentane,  the experimentally measured data  agrees with using $r=2$ with a margin of 20 percent \cite{Zhao2021}. 

In the case of glasses,  Cahill, Watson and Pohl \cite{Cahill_1992}  derived a model of the minimum of thermal conductivity following an idea originally proposed by Einstein according to which thermal energy is transported by harmonic interactions between vibrating atoms with random phases. The derived expression for amorphous solids in the high field temperature limit becomes \cite{Xie_2017}:

\begin{equation}
\kappa_G=\left(\frac{\pi}{48}\right)^{1/3}  n^{2/3} k_{\rm B} (v_l+ 2v_t)
\label{Cahill}
\end{equation}

Assuming equality between longitudinal, $v_l$,  and transverse, $v_t$, phonon velocities, that is $v_l=v_t=v_s$, this equation becomes identical  to the Bridgman formula, albeit with $r=1.21$ \cite{Xi_2020,Zhao2021}.

Like the Bridgman formula \cite{Brigman1923} and the asymptotic case of the Cahill-Pohl equation \cite{Cahill_1992,Xie_2017}, our equation 4 includes $n^{2/3}$ and $v_{th}=\sqrt{\frac{2 k_B T}{m}}= \frac{v_F}{\sqrt{\pi}}\frac{\lambda_F}{\Lambda}$ rather than the sound velocity. An insight to this difference  is provided by comparing the entropy production rate, which is proportional to the thermal conductivity at constant temperature and temperature gradient. In a classical system, the time scale for entropy production rate is set by the ratio of the interatomic distance divided by the sound velocity. In contrast, in a Fermi-Dirac distribution, the fundamental time scale is the ratio of the Fermi velocity to the de Broglie thermal length. As a consequence, the relevant group velocity becomes temperature dependent.

Note that in a Fermi liquid, the sound velocity v$_s$ and the Fermi velocity $v_F$ are intimately linked \cite{Vollhardt1990}:

\begin{equation}
v_s^2=\frac{1}{3}(1+F^s_0)(1+\frac{1}{3}F^s_1)v_F^2
\label{vs-vf}
\end{equation}

Here, $F^s_0$  and $F^s_1$ are Landau parameters. When $\Lambda_F=\Lambda$ at the quantum/classical crossover, we have  v$_{th}\approx$ v$_s$ and our equation   becomes yet another version of the classical Bridgman formula. 
\end{widetext}
\end{document}